\documentclass[]{piparticle-final}
\usepackage{graphicx}
\usepackage{amsmath,amssymb,amsthm,xspace,enumerate,graphicx,multirow,url}
\usepackage{multirow}
\usepackage[utf8]{inputenc} 
\usepackage{pgfplots}
\pgfplotsset{compat=newest}

\usepackage{cite} 
\usepackage{epstopdf}

\newcommand{\NN}{\mathbb{N}}
\newcommand{\RR}{\mathbb{R}}
\newcommand{\concat}{{}^\smallfrown}
\newcommand{\K}{K_{\M}}
\newcommand{\M}{N}
\newcommand{\words}{\{0,1\}^*}
\newcommand{\leng}{{\rm LT^2C^2}\xspace}
\newcommand{\xsimple}{x_2}
\newcommand{\xpi}{x_3}
\newcommand{\xrandom}{x_1}
\newcommand{\eqdef}                     
  {\stackrel{\scriptscriptstyle \mathrm{def}}{=}}

\theoremstyle{plain}

\theoremstyle{remark}

\begin{document}

\volume{5}               
\articlenumber{050001}   
\journalyear{2013}       
\editor{G. Mindlin}   
\received{12 December 2012}     
\accepted{2 February 2013}   
\runningauthor{S Romano \itshape{et al.}}  
\doi{050001}         

\title{$\leng$: A language of thought with Turing-computable Kolmogorov complexity}

\author{    Sergio Romano,\cite{dc}\thanks{E-mail: sgromano@dc.uba.ar}\hspace{0.5em} 
            Mariano Sigman,\cite{df,conicet}\thanks{E-mail: sigman@df.uba.ar}\hspace{0.5em} 
            Santiago Figueira\cite{dc,conicet}\thanks{E-mail: santiago@dc.uba.ar}
            }

\pipabstract{In this paper, we present a theoretical effort to  connect the theory
of program size to psychology by implementing  a concrete  language
of  thought with Turing-computable Kolmogorov complexity ($\leng$)
satisfying the following requirements: 1)  to be simple enough so
that the complexity of any given finite binary sequence can be
computed, 2)  to be based on tangible operations of human reasoning
(\textsl{printing}, \textsl{repeating},\dots), 3)  to be sufficiently
powerful to generate all possible sequences but not too powerful as
to identify regularities which would be {\em invisible} to humans.
We first formalize $\leng$, giving its syntax and semantics and
defining an adequate notion of {\em program size}. Our setting leads
to a Kolmogorov complexity function relative to $\leng$ which is
computable in polynomial time, and it also induces a prediction
algorithm in the spirit of Solomonoff's inductive inference theory.
We then prove the efficacy of this language by investigating
regularities in strings produced by participants attempting to
generate random strings. Participants had a profound understanding
of randomness and hence avoided typical misconceptions such as
exaggerating the number of alternations. We reasoned that remaining
regularities would express the algorithmic nature of human thoughts,
revealed in the form of specific patterns. Kolmogorov complexity
relative to $\leng$ passed three expected tests examined here: 1)
human sequences were less complex than control PRNG sequences, 2)
human sequences were not stationary, showing decreasing values of
complexity resulting from fatigue, 3) each individual showed traces
of algorithmic stability since fitting of partial sequences was more
effective to predict subsequent sequences than average fits. This work
extends on previous efforts to combine notions of Kolmogorov
complexity theory and algorithmic information theory to psychology,
by explicitly proposing a language which may describe the patterns
of human thoughts.
}

\maketitle

\blfootnote{
\begin{theaffiliation}{99}
   \institution{dc} Department of Computer Science, FCEN, University of Buenos Aires, Pabellón I, Ciudad Universitaria (C1428EGA) Buenos Aires, Argentina. 
   \institution{df} Laboratory of Integrative Neuroscience, Physics Department, FCEN, University of Buenos Aires, Pabellón I, Ciudad Universitaria (C1428EGA) Buenos Aires, Argentina.
   \institution{conicet} CONICET, Argentina.
\end{theaffiliation}
}

\section{Introduction}
Although people feel they understand the concept of
randomness~\cite{kac83}, humans are unable to produce random
sequences, even when instructed to do so~\cite{Reichenbach1949,
tune64, Baddeley96, TverskyKahneman71, Wagenaar91}, and to perceive
randomness in a way that is inconsistent with probability theory
~\cite{falk75, falk81,kahnemant72, KahnemanTversky72}. For instance,
random sequences are not perceived by participants as such because
runs 
appear too long to be
random~\cite{Gilovich85, Wgenaar88} and, similarly, sequences
produced by participants aiming to be random have too many
alternations ~\cite{Budescu, Rapoport}. This bias, known as the
gambler's fallacy, is thought to result from an expectation of local
representativeness (LR) of
randomness~\cite{KahnemanTversky72} which ascribes chance to a
self-correcting mechanism, promptly restoring the balance whenever
disrupted. In words of Tversky and Kahneman~\cite{TverskyKahneman71},
people apply the law of large numbers too hastily, as if it were the
\textit{law of small numbers}. The gambler's fallacy  leads to
classic psychological illusions in real-world situations such as the
\textit{hot hand} perception by which people assume specific states
of high performance, while analysis of records show that sequences
of hits and misses are largely compatible with Bernoulli (random)
process~\cite{LSK89,TG89}.

Despite massive evidence showing that perception and productions of
randomness shows systematic distortions, a mathematical and
psychological theory of randomness remains partly elusive. From a
mathematical point of view ---as discussed below--- a notion of
randomness for finite sequences presents a major challenge.

From a psychological point of view, it remains difficult to ascribe
whether the inability to produce and perceive randomness adequately
results from a genuine misunderstanding of randomness or, instead,
as a consequence of the algorithmic nature of human thoughts which is  revealed in the forms of patterns and, hence, in the
impossibility of producing genuine chance.

In this work, we address both issues by developing  a framework based on a specific \textit{language of
thought} by instantiating a simple device which induces a computable
(and efficient) definition of algorithmic
complexity~\cite{K65,C75,LZ70}.

The notion of algorithmic complexity is described in greater
detail below but, in short, it assigns a measure of complexity to a given
sequence as the length of the shortest program capable of producing
it. If a sequence is algorithmically compressible, it implies that
there may be a certain pattern embedded (described succinctly by the
program) and hence it is not random. For instance, the binary
version of Champernowne's sequence~\cite{C33}
\begin{center}
\small$0 1 10 11 100 101 110 111 1000 1001 1010 1011 1100 \dots$
\end{center} consisting of the concatenation of the binary representation of all the natural numbers, one after another, is known to be {\em normal} in
the scale of 2, which means that every finite word of length $n$
occurs with a limit frequency of $2^{-n}$ ---e.g.,\ the string $1$
occurs with probability $2^{-1}$, the string $10$ with probability
$2^{-2}$, and so on. Although this sequence may seem random based on
its probability distribution, every prefix of length $n$ is produced
by a program much shorter than $n$.

The theory of program size, developed simultaneously in the '60s by
Kolmogorov~\cite{K65}, Solomonoff~\cite{So64} and Chaitin~\cite{C69},
had a major influence in theoretical computer science.  Its
practical relevance was rather obscure because most notions, tools
and problems were undecidable and, overall, because it did not apply
to finite sequences. A problem at the heart of this theory is
that the complexity of any given sequence depends on the chosen language. 
For instance, the sequence
\begin{center}
\small $\xrandom=1100101001111000101000110101100110011100$
\end{center}
which seems highly complex, may be trivially accounted by a single
character if there is a symbol (or instruction of a programming
language) which accounts for this sequence. This has its
psychological analog in the kind of regularities people often
extract:
\begin{center}
\small $\xsimple=1010101010101010101010101010101010101010$
\end{center}
is obviously a non-random sequence, as it can succinctly be
expressed as

\begin{equation}\label{eqn:short-program}
\small \textsl{repeat 20 times: print `10'}.
\end{equation}
Instead, the sequence
\begin{center}
\small $\xpi=0010010000111111011010101000100010000101$
\end{center}
appears more random and yet it is highly compressible as it  consists
of the first 40 binary digits of $\pi$ after the decimal point.
This regularity is simply not extracted by the
\textit{human-compressor} and demonstrates how the exceptions
to randomness reveal natural patterns of thoughts~\cite{Griff03}.

The genesis of a practical (computable) algorithmic information
theory~\cite{LV08} has had an influence (although not yet a major
impact) in psychology. Variants of Kolmogorov complexity have been
applied to  human concept learning~\cite{Feldman00}, to general
theories of cognition~\cite{Chater99} and to subjective
randomness~\cite{falkk97, Griff03}. In this last work, Falk and
Konold showed that a simple measure, inspired in algorithmic notions, was a good correlate of perceived randomness~\cite{falkk97}.
Griffiths \& Tenenbaum developed statistical models that incorporate
the detection of certain regularities, which are classified in terms
of the Chomsky hierarchy~\cite{Griff03}. They showed the existence
of motifs (repetition, symmetry) and related their probability
distributions to Kolmogorov complexity via Levin's coding theorem
(cf.~section \ref{sec:discussion} for more details).

The main novelty of our work is to develop a class of specific
programming languages (or Turing machines) which allows us to stick
to the theory of program size developed by Kolomogorov, Solomonoff
and Chaitin. We use the patterns of sequences of humans aiming to
produce random strings to fit, for each individual, the language which
 captures these regularities.

\section{Mathematical theory of randomness}\label{sec:intro}
%
%


The idea behind Kolmogorov complexity theory is to study the length
of the descriptions that a formal language can produce to identify a
given string. All descriptions are finite words over a finite
alphabet, and hence each description has a finite length
---or, more generally, a suitable notion of {\em size}.
One string may have many descriptions, but any description should
describe one and only one string. Roughly, the Kolmogorov
complexity~\cite{K65} of a string $x$ is the length of the shortest
description of $x$. So a string is `simple' if it has at least one
short description, and it is `complex' if all its descriptions are
long. Random strings are those with high complexity.

As we have mentioned, Kolmogorov complexity uses programming languages to
describe strings. Some programming languages are Turing complete,
which means that any partial computable function can be represented
in it. The commonly used programming languages, like C++ or Java,
are all Turing complete. However, there are also Turing incomplete
programming languages, which are less powerful but more convenient
for specific tasks.

In any reasonable imperative language, one can describe $\xsimple$
above with a program like (\ref{eqn:short-program}), %
of length 26, which is considerably smaller than $40$, the size of
the described string. It is clear that $\xsimple$ is `simple'.
%
The case of $\xpi$ is a bit tricky. Although at first sight it seems
to have a complete lack of structure, it contains a hidden pattern:
it consists of the first forty binary digits of $\pi$ after the
decimal point. This pattern could hardly be recognized by the
reader, but once it is revealed to us, we agree that $\xpi$ must
also be tagged as `simple'. Observe that the underlying programming
language is central: $\xpi$ is `simple' with the proviso that the
language is strong enough to represent (in a reasonable way) an
algorithm for computing the bits of $\pi$
---a language to which humans are not likely to have access when they try to find patterns in a string.
Finally, for $\xrandom$, the best way to describe it seems to be
something like
\begin{center}
\small \textsl{print `1100101001111000101000110101100110011100'},
\end{center}
which includes the string in question verbatim,  length $48$.
Hence $\xrandom$ only has long descriptions and hence it is
`complex'.

In general, both the string of length $n$ which alternates $0$s and
$1$s and the string which consists of the first $n$ binary digits
of $\pi$ after the decimal point can be computed by a program of
length $\approx \log n$ ---and this applies to any computable
sequence. The idea of the algorithmic randomness theory is that a
truly random string of length $n$ necessarily needs a program of
length $\approx n$ (cf.\ section \ref{sec:randomness_finite} for details).

\subsection{Languages, Turing machines and Kolmogorov complexity}\label{sec:kolmogorov}

Any programming language $\+L$ can be formalized with a Turing
machine $M_\+L$, so that programs of $\+L$ are represented as inputs
of $M_\+L$ via an adequate binary codification. If $\+L$ is Turing
complete then the corresponding machine $M_\+L$ is called {\em
universal}, which is equivalent to say that $M_\+L$ can simulate any
other Turing machine.

%

Let $\words$ denote the set of finite words over the binary
alphabet. Given a Turing machine $M$, a program $p$ and a string $x$
($p,x\in\words$), we say that $p$ is an $M$-description of $x$ if
$M(p)=x$ ---i.e.,\ the program $p$, when executed in the machine $M$,
computes $x$. Here we do not care about the time that the
computation needs, or the memory it consumes. The Kolmogorov
complexity of $x\in\words$ relative to $M$ is defined by the length of the shorter $M$-description of $x$. More formally, \begin{eqnarray*}
&K_M(x)\eqdef\min\{|p| \colon M(p)=x\}\cup\{\infty\},&
\end{eqnarray*}
where $|p|$ denotes the length of $p$. Here $M$ is {\em any} given Turing machine, possibly one with a very specific behavior, so it may be
the case that a given string $x$ does not have any $M$-description
at all. In this case, $M(x)=\infty$.  In practical terms, a machine
$M$ is a useful candidate to measure complexity if it  computes a
surjective function. In this case, every string $x$ has at least one
$M$-description and therefore $K_M(x)<\infty$.

\subsection{Randomness for finite words}\label{sec:randomness_finite}

The strength of Kolmogorov complexity appears when $M$ is set to any
universal Turing machine $U$. The invariance theorem states that
$K_U$ is minimal, in the sense that for every Turing machine $M$
there is a constant $c_M$ such that for all $x\in\words$ we have
$K_U(x)\leq K_M(c)+c_M$. Here, $c_M$ can be seen as the specification of the language $M$ in $U$ (i.e.,\ the information contained in $c_M$ tells $U$ that the machine to be simulated is $M$). If $U$ and $U'$ are two universal Turing
then $K_U$ and $K_{U'}$ differ at most by a constant. In a few
words, $K_U(x)$ represents the length of the ultimate compressed
version of $x$, performed by means of algorithmic processes.

For analysis of arbitrarily long sequences, $c_M$ becomes negligible and hence for nonpractical aspects of the theory the choice of the machine is not relevant. However, for short sequences, as we study
here, this becomes a fundamental problem, as notions of
complexity are highly dependent on the choice of the underlying machine through
the constant $c_M$. The most trivial example, as referred in the
introduction, is that for any given sequence, say $x_1$, there is a machine $M$
for which $x_1$ has minimal complexity.

\subsection{Solomonoff induction}\label{sec:solomonoff}

Here we have presented {\em compression} as a framework to
understand randomness. Another very influential paradigm proposed by
Schnorr is to use the notion of {\em martingale} (roughly, a betting
strategy), by which a sequence is random if there is no computable
martingale capable of predicting forthcoming symbols (say, of a
binary alphabet $\{0,1\}$) better than chance~\cite{S71,S71b}. In
the 1960s, Solomonoff~\cite{So64} proposed a universal prediction
method which successfully approximates any distribution $\mu$, with
the only requirement of $\mu$ being computable.

This theory brings together concepts of algorithmic information,
Kolmogorov complexity and probability theory. Roughly, the idea is
that amongst all `explanations' of $x$, those which are `simple' are
more relevant, hence following Occam's razor principle: amongst all
hypothesis that are consistent with the data, choose the simplest.
Here the `explanations' are formalized as programs computing $x$,
and `simple' means low Kolmogorov complexity.

Solomonoff's theory, builds on the notion of {\em monotone} (and
prefix) Turing machines. Monotone machines are ordinary Turing
machines with a one-way read-only input tape, some work tapes, and a
one-way {\em write-only} output tape. The output is written one
symbol at a time, and no erasing is possible in it. The output can
be finite if the machine halts, or infinite in case the machine
computes forever. The output head of monotone machines can only
``print and move to the right"  so they are  well suited for the
problem of inference of forthcoming symbols based on partial (and
finite) states of the output sequence. Any monotone machine $N$ has
the {\em monotonicity} property (hence its name) with respect to
extension: if $p,q\in\words$ then $\M(p)$ is a prefix of
$\M(p\concat q)$, where $p\concat q$ denotes the concatenation of
$p$ and $q$.

One of Solomonoff's fundamental results is that given a finite
observed sequence $x\in\words$, the most likely finite continuation
is the one  in which the concatenation of $x$ and $y$ is less
complex in a Kolmogorov sense. This is formalized in the following
result (see theorem\ 5.2.3 of~\cite{LV08}): for almost all infinite
binary sequences $X$ (in the sense of $\mu$) we have
\begin{center}
$\small\small -\lim\limits_{n\to\infty}\log\mu(y\ |\ X\!\!
\upharpoonright \! n)=$

$\small\small\lim\limits_{n\to\infty}Km_U((X \!\!  \upharpoonright
\! n)\concat y)-Km_U(X \!\!  \upharpoonright \!  n)+O(1)<\infty.$
\end{center}
Here, $X\! \! \upharpoonright \! n$ represents the first $n$ symbols
of $X$, and $Km_U$ is the monotone Kolmogorov complexity relative to
a monotone universal machine $U$. That is, $Km_U(x)$ is defined as
the length of the shortest program $p$ such that the output of
$U(p)$ starts with $x$
---and possibly has a (finite or infinite) continuation.

In other words, Solomonoff inductive inference leads to a method of
prediction based on data compression, whose idea is that whenever
the source has output the string $x$, it is a good heuristic to
choose the extrapolation $y$ of $x$ that minimizes $Km_U(x\concat
y)$. For instance, if one has observed $\xsimple$, it is more likely
 for the continuation to be $1010$ rather than $0101$, as the former
can be succinctly described by a program like
\begin{eqnarray}\label{eqn:continuation1}
&&\small\textsl{repeat 22 times: print `10'}.
\end{eqnarray}
and the latter looks more difficult to describe; indeed the shorter
program describing it seems to be something like
\begin{eqnarray}\label{eqn:continuation2}
&&\small\textsl{repeat 20 times: print `10';}\\
&&\small\textsl{print `0101'}.\nonumber
\end{eqnarray}
Intuitively, as  program (\ref{eqn:continuation1}) is shorter than
(\ref{eqn:continuation2}), $\xsimple\concat 1010$ is more probable
than $\xsimple\concat 0101$. Hence, if we have seen $\xsimple$, it seems to be  a better strategy to
predict $1$.

\section{A framework for human thoughts}\label{sec:framework}
The notion of thought is not well grounded.  We lack an operative
working definition and, as also happens with other terms in
neuroscience (consciousness, self, ...), the word \textit{thought} is
highly polysemic in common language. It may refer, for example, to a
belief, to an idea or to the content of the conscious mind. Due to
this difficulty, the mere notion of \textit{thought} has not been a
principal or directed object of study in neuroscience, although of
course it is always present implicitly, vaguely, without a formal
definition.

Here we do not intend to elaborate an extensive review on the
philosophical and biological conceptions of thoughts (see
\cite{Dellarosa} for a good review on thoughts). Nor  are we in a
theoretical position to provide a full formal definition of a
thought. Instead, we point to the key assumptions of our framework
about the nature of thoughts. This accounts to defining constraints
in the class of thoughts which we aim to describe. In other words,
we do not claim to provide a general theory of human thoughts (which
is not amenable at this stage lacking a full definition of the
class) but rather of a subset of thoughts which satisfy certain
constraints defined below.

For instance, E.B.\ Titchener and W.\ Wundt, the founders of
 structuralist school in psychology (seeking structure in the mind
without evoking metaphysical conceptions, a tradition which we
inherit and to which we adhere), believed that thoughts were images
(there are not \textit{imageless thoughts}) and hence can be broken
down to elementary sensations \cite{Dellarosa}. While we do not
necessarily agree with this propositions (see Carey
\cite{carey2009origin} for more contemporary versions denying the
sensory foundations of conceptual knowledge), here we do not intend
to explain all possible thoughts but rather a subset, a simpler
class which ---in agreement with the Wundt and Titchener--- can be
expressed in images. More precisely, we develop a theory which may
account for Boole's \cite{boole1854investigation}
 notion of thoughts as propositions and statements about the world
which can be represented symbolically.  Hence, a first and crucial
assumption of our framework is that thoughts are discrete. Elsewhere
we have extensively discussed
\cite{zylberberg2009neurophysiological,zylberberg2011human,graziano2011parsing,zylberberg2012construction,shalom2011looking,dehaene2012single,kamienkowski2011effects}
how the human brain,  whose architecture is quite different from
Turing machines, can emerge in a form of computation which is
discrete, symbolic and resembles Turing devices.

Second, here we focus on the notion of ``prop-less'' mental
activity, i.e.,\ whatever (symbolic) computations can be carried out
by  humans without resorting to external aids such as paper, marbles,
computers or books. This is done by actually asking  participants to
perform the task ``in their heads''.  Again, this is not intended to
set a proposition about the universality of human thoughts but,
instead, a narrower set of thoughts which we conceive is theoretically addressable
 in this mathematical framework.

Summarizing:
\begin{enumerate}
\item  We think we do not have  a good mathematical (even philosophical) conception of thoughts, as mental structures, yet.
\item  Intuitively (and philosophically), we adhere to a materialistic and computable approach to thoughts. Broadly, one can think (to picture, not to provide a formal framework) that thoughts are formations of the mind with certain stability which defines distinguishable clusters or objects \cite{zylberberg2010brain,gallos2012small,gallos2012conundrum}.
\item  While the set of such objects and the rules of their transitions may be of many different forms (analogous, parallel, unconscious, unlinked to sensory experience, non-linguistic, non-symbolic), here we work on a subset of thoughts, a class defined by Boole's attempt to formalize thought as symbolic propositions about the world.
\item  This states ---which may correspond to human ``conscious rational thoughts'', the seed of Boole and Turing foundations \cite{zylberberg2011human,zylberberg2011human}--- are discrete and defined by symbols and potentially represented by a Turing device.
\item  We focus on an even narrower space of thoughts. Binary formations (right or left, zero or one)  to focus on what kind of language better describes these transitions. This work can be naturally extended to understand discrete transitions in conceptual formations \cite{costa2009scale,mota2012speech,sigman2002global}.
\item  We concentrate on prop-less mental activity to understand limitations of the  human mind when it does not have evident external support (paper, computer...)
\end{enumerate}

\section{Implementing a  language of  thought with Turing-computable complexity}\label{sec:model}
As explained in section \ref{sec:intro}\ref{sec:kolmogorov}, Kolmogorov
complexity considers all possible computable compressors and assigns
to a string $x$ the length of the shortest of the corresponding compressions. This
seems to be a perfect theory of compression but it has a drawback:
the function $K_U$ is not computable, that is, there is no effective
procedure to calculate $K_U(x)$ given $x$.

On the other hand, the definition of randomness introduced in
section \ref{sec:intro}\ref{sec:kolmogorov}, having very deep and
intricate connections with algorithmic information and computability
theories, is simply too strong to explain our own perception of
randomness. To detect that $\xpi$ consists of the first twenty bits
of $\pi$ is incompatible with human patterns of thought.

Hence, the intrinsic algorithms (or observed patterns) which make
human sequences not random are too restricted to be accounted by a
{\em universal} machine and may be better described by a {\em
specific} machine.  Furthermore, our hypothesis is that each person
uses his own particular specific machine or algorithm to generate a
random string.

As a first step in this complicated enterprise, we propose to work
with a specific language $\leng$ which meets the following
requirements:
\begin{itemize}
\item $\leng$ must reflect some plausible features of our mental activity
when finding succinct descriptions of words. For instance, finding
repetitions in a sequence such as $\xsimple$ seems to be something
easy for our brain, but detecting numerical dependencies between its
digits as in $\xpi$ seems to be very unlikely.

\item $\leng$ must be able to describe any string in $\words$. This means that the
map given by the induced machine $\M\eqdef\M_{\leng}$ must be
surjective.

\item $\M$ must be simple enough so
that $K_{\M}$ ---the Kolmogorov complexity relative to $\M$---
becomes computable. This requirement clearly makes $\leng$ Turing
incomplete, but as we have seen before, this is consistent with
human deviations from randomness.

\item The rate of compression given by $K_{\M}$ must be sensible for
very short strings, since our experiments will produce such strings.
For instance, the approach, followed in \cite{CV05}, of using the size
of the compressed file via general-purpose compressors like
Lempel-Ziv based dictionary (gzip) or block based (bzip2) to
approximate the Kolmogorov complexity does not work in our setting.
This method works best for long files.

\item $\leng$ should have certain degrees of freedom, which can be adjusted in order to approximate
the specific machine that each individual follows during the process
of randomness generation.
\end{itemize}

We will not go into the details on how to codify the instructions of
$\leng$ into binary strings of $\M$: for the sake of simplicity we
take $\M$ as a surjective total mapping $\leng\to \words$.  We
restrict ourselves to describe the grammar and semantics of our
proposed programming language $\leng$. It is basically an imperative
language with only two classes of instructions: a sort of
\textsl{print $i$}, which prints the bit $i$ in the output; and a
sort of \textsl{repeat $n$ times $P$}, which for a fixed $n\in\NN$
it repeats $n$ times the program $P$. The former is simply
represented as $i$ and the latter as $(P)^n$.

Formally, we set the alphabet $\{0,1,(,),^0,\dots,^9\}$ and define
$\leng$ over such alphabet with the following grammar:

$$
P\quad ::= \quad\epsilon\quad|\quad 0\quad | \quad 1 \quad |\quad PP
\quad | \quad (P)^n,
$$
where $n>1$ is the decimal representation of $n\in\NN$ and
$\epsilon$ denotes the empty string. The semantics of $\leng$ is
given through the behavior of $\M$ as follows:
\begin{eqnarray*}
\M(i)&\eqdef&p\qquad \mbox{for $i\in\{\epsilon,0,1\}$}\\
\M(P_1P_2)&\eqdef&\M(P_1)\concat \M(P_2)\\
\M((P)^n)&\eqdef&\underbrace{\M(P)\concat\cdots\concat
\M(P)}_{\mbox{\small $n$ times}}.
\end{eqnarray*}
$\M$ is not universal, but every string $x$ has a program in $\M$
which describes it: namely $x$ itself. Furthermore, $\M$ is monotone
in the sense that if $p,q\in\leng$ then $\M(p)$ is a prefix of
$\M(p\concat q)$.
In Table \ref{tab:example}, the first column shows some examples of
$\M$-programs which compute $1001001001$.

\begin{table}[h]
\centering
\begin{tabular}{rr}
{\em program}                 & {\em size}    \\
\hline
\hline
$1001001001$        & 10    \\
$(100)^21(0)^21$    & 6.6     \\ 
$(100)^31$          & 4.5    \\ 
$1((0)^21)^3$       & 3.8 \\
\hline
\hline
\end{tabular}
\caption{Some $\M$-descriptions of $1001001001$ and its sizes for
$b=r=1$} \label{tab:example}
\end{table}

\subsection{Kolmogorov complexity for $\leng$}\label{subsec:KL}

The Kolmogorov complexity relative to $\M$ (and hence to the
language $\leng$) is defined as

$$
K_{\M}(x)\eqdef\min\{\|p\| \colon p\in\leng, \M(p)=x\},
$$
where $\|p\|$, the {\em size} of a program $p$, is inductively
defined as:

\begin{eqnarray*}
\|\epsilon\|&\eqdef&0\\
\|p\|&\eqdef&b\qquad \mbox{for $p\in\{0,1\}$}\\
\|P_1P_2\|&\eqdef&\|P_1\|+\|P_2\|\\
\|(P)^n\|&\eqdef&r\cdot\log n+\|P\|.
\end{eqnarray*}
In the above definition, $b \in \NN , r \in \RR$ are two parameters
that control the relative weight of the {\em print} operation and
the {\em repeat $n$ times} operation. In the sequel, we drop the
subindex of $\K$ and simply write $K\eqdef\K$.
Table~\ref{tab:example} shows some examples of the size of
$\M$-programs when $b=r=1$. Observe that for all $x$ we have
$K(x)\leq\|x\|$.

It is not difficult to see that $K(x)$ depends only on the values of
$K(y)$, where $y$ is any nonempty and proper substring of $x$. Since
$\|\cdot\|$ is computable in polynomial time, using dynamic
programming one can calculate $K(x)$ in polynomial time. This, of
course, is a major difference with respect to the Kolmogorov
complexity relative to a {\em universal} machine, which is not
computable.

\subsection{From compression to prediction}\label{subsec:predictor}

As one can imagine, the perfect universal prediction method
described in section \ref{sec:intro}\ref{sec:solomonoff} is, again,
non-computable. We define a computable prediction algorithm  based
on Solomonoff's theory of inductive inference but using $K$, the
Kolmogorov complexity relative to $\leng$, instead of $Km_U$ (which
depends on a universal machine). To predict the next symbol of
$x\in\words$, we follow the idea described in
section \ref{sec:intro}\ref{sec:solomonoff}: amongst all extrapolations
$y$ of $x$ we choose the one that minimizes $K(x\concat y)$. If such
$y$ starts with $1$, we predict $1$, else we predict $0$. Since we
cannot examine the infinitely many extrapolations, we restrict to
those  up to a fixed given length~$\ell_F$. Also, we do not take
into account the whole $x$ but only a suffix of length $\ell_P$.
Both $\ell_F$ and $\ell_P$ are parameters which control,
respectively, how many extrapolation bits are examined ($\ell_F$
many {\em Future} bits) and how many bits of the tail of $x$
($\ell_P$ many {\em Past} bits) are considered.

Let $\{0,1\}^n$ (resp.\ $\{0,1\}^{\leq n}$) be the set of words over
the binary alphabet $\{0,1\}$ of length $n$ (resp.\ at most $n$).
Formally, the prediction method is as follows. Suppose $x=x_1\cdots
x_n$ ($x_i\in \{0,1\}$) is a string. The next symbol is determined
as follows:

$$\small
{\rm Next}(x_1 \cdots x_n)\eqdef
\begin{cases}
0&\mbox{if $m_0<m_1$;}\\
1&\mbox{if $m_0>m_1$;}\\
g(x_{n-\ell_P}\cdots x_n)&\mbox{otherwise.}
\end{cases}
$$
where for $i\in\{0,1\}$,

$$
m_i\eqdef\min\{K(x_{n-\ell_P}\cdots x_n i\concat y)\colon
y\in\{0,1\}^{\leq \ell_F}\},
$$
and $g:\{0,1\}^{\ell_P}\to\{0,1\}$ is defined as $g(z)=i$ if the
number of occurrences of $i$ in $z$ is greater than the number of
occurrences of $1-i$ in $z$; in case the number of occurrences of
$1$s and $0$s in $z$ coincide then $g(z)$ is defined as the last bit
of $z$.

\section{Methods}\label{sec:experiment}
Thirty eight volunteers (mean age = 24) participated in an
experiment to examine the capacity of  $\leng$ to identify
regularities in production of binary sequences. Participants were
asked to produce random sequences, without further instruction.

All the participants were college students or graduates with
programming experience and knowledge of the theoretical
foundations of randomness and computability. This was intended
to test these ideas in a  {\em hard} sample where we did not expect
typical errors which results from a misunderstanding of chance.

The experiment was divided in four blocks. In each block the
participant pressed freely the left or right arrow $120$ times.

After each key press, the participant received a notification with a
green square which progressively filled a line to indicate the
participant the number of choices made. At the end of the block,
participants were provided feedback of how many times the predictor
method has correctly predicted their input. After this point, a new
trial would start.

$38$ participants performed 4 sequences, yielding a total of $152$
sequences. $14$ sequences were excluded from analysis because they
had an extremely high level of predictability. Including these
sequences would have actually improved all the scores reported here.

The experiment was programmed in ActionScript and can be seen
at~\url{http://gamesdata.lafhis-server.exp.dc.uba.ar/azarexp}.

\section{Results}\label{sec:results}
\subsection{Law of large numbers}

Any reasonable notion of randomness for strings on base $2$ should
imply Borel's normality, or the law of large numbers in the sense
that if $x\in \{0,1\}^n$ is random then the number of occurrences of
any given string $y$ in $x$ divided by $n$ should tend to $2^{-|y|}$,
as $n$ goes to infinity.

A well-known result obtained in some investigations on generation or
perception of randomness in binary sequences is that people tend to
increase the number of alternations of symbols with respect to the
expected value \cite{falkk97}. Given a string $x$ of length $n$ with
$r$ runs, there are $n - 1$ transitions between successive symbols
and the number of alternations between symbol types is $r - 1$. The
\textit{probability of alternation} of the string $x$ is defined as
\begin{eqnarray*}
&P(x):\{0,1\}^{\geq 2} \to [0,1]&\\
&P(x)=\frac{r-1}{n-1}.&
\end{eqnarray*}
%
In our experiment, the average
$P(x)$ of  participants was $0.51$,
very close to the expected \textit{probability of alternation} of a
random sequence which should be $0.5$. A t-test on the $P(x)$ of the
strings produced by participants, where the null hypothesis is that
they are a random sample from a normal distribution with mean $0.5$,
shows that the hypothesis cannot be rejected as the $p$-value is
$0.31$ and the confidence interval on the mean is $[0.49,0.53]$.
This means that the \textit{probability of alternation} is not a
good measure to distinguish participant's strings from random ones,
or at least, that the participants in this very experiment can
bypass this validation.

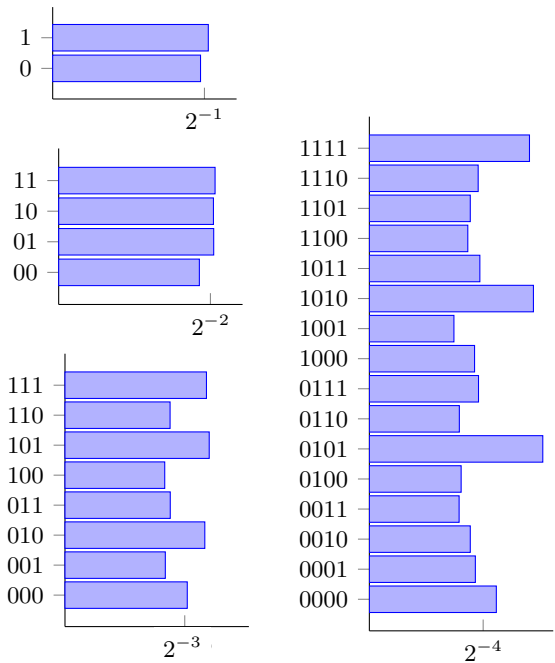
\begin{figure}[htb]
\begin{center}
\begin{tabular}[t]{cc}
\pgfplotsset{width=4cm, height=2.8cm}
\begin{tikzpicture}
  \tikzstyle{every node}=[font=\small] \centering
  \begin{axis}[
    xbar,
    enlargelimits=1.0,
    enlarge x limits={upper,value=0.01},
    symbolic y coords={0,1},
    ytick=data,
    xmin=0,xmax=.6,
    axis lines*=left,
    xtick={0.5},
    ]
    \addplot coordinates {(0.487198068,0) (0.512801932,1) };
  \end{axis}
\coordinate (2 on x)   at (2,-.01); \node [rectangle,fill=white,below]  at  (2 on x) {\small $2^{-1}$};
\end{tikzpicture}
& \multirow{3}{*}{\pgfplotsset{width=4cm, height=8.4cm}
\begin{tikzpicture}
 \tikzstyle{every node}=[font=\small]
%
  \centering
  \begin{axis}[
    xbar=5,
    enlargelimits=.07,
    enlarge x limits={upper,value=0.01},
    symbolic y coords={0000, 0001, 0010, 0011, 0100, 0101, 0110, 0111, 1000, 1001, 1010, 1011, 1100, 1101, 1110, 1111},
    ytick=data,
    xmin=0,xmax=.1,
    axis lines*=left,
    xtick={0.0625},
   x tick label style={
        /pgf/number format/.cd,
        fixed,
        fixed zerofill,
        precision=4,
        /tikz/.cd
        },
    ]
    \addplot coordinates {(0.06973863,0000)
(0.05821875,0001) (0.05543169,0010) (0.04930014,0011) (0.05041496
,0100) (0.09525579,0101) (0.04942401,0110) (0.05995293,0111)
(0.05778521,1000) (0.04645113,1001) (0.0901152,1010)
(0.06069615,1011) (0.05406912,1100) (0.05543169,1101)
(0.05976712,1110) (0.08794748,1111)
 };
  \end{axis}
\coordinate (1 on x)   at (1.5,-.01); \node [rectangle,fill=white,below]  at  (1 on x) {\small $\ \ 2^{-4}\ \ $};
\end{tikzpicture}}
\\
\pgfplotsset{width=4cm, height=3.65cm}
\begin{tikzpicture}
 \tikzstyle{every node}=[font=\small]
   \centering
  \begin{axis}[
    xbar=5,
    enlargelimits=.35,
    enlarge x limits={upper,value=0.01},
    symbolic y coords={00,01,10,11},
    ytick=data,
    xmin=0,xmax=.3,
    axis lines*=left,
    xtick={0.25},
    ]
    \addplot coordinates {(0.231823164,00)
(0.255450006,01) (0.255023749,10) (0.257703081,11) };
  \end{axis}
  \coordinate (2 on x)   at (2,-.01); \node [rectangle,fill=white,below]  at  (2 on x) {\small $2^{-2}$};
\end{tikzpicture}
&
\\\pgfplotsset{width=4cm, height=5.2cm}
\begin{tikzpicture}
 \tikzstyle{every node}=[font=\small]
  \centering
  \begin{axis}[
    xbar=5,
    enlargelimits=.15,
    enlarge x limits={upper,value=0.01},
    symbolic y coords={000,001,010,011,100,101,110,111},
    ytick=data,
    xmin=0,xmax=.19,
    axis lines*=left,
    xtick={0.125},
   x tick label style={
        /pgf/number format/.cd,
        fixed,
        fixed zerofill,
        precision=3,
        /tikz/.cd
        },
    ]
    \addplot coordinates {(0.127609924,000)
(0.104704004,001) (0.145910096,010) (0.109862442,011)
(0.104089904,100) (0.150515844,101) (0.109678212,110)
(0.147629575,111)
 };
  \end{axis}
    \coordinate (2 on x)   at (1.55,-.01); \node [rectangle,fill=white,below]  at  (2 on x) {\small $2^{-3}$};
\end{tikzpicture}
&
\end{tabular}
\end{center} \caption{Frequency of sub-strings up to
length 4} \label{fig:largeNumbersLaws}
\end{figure}

Although the \textit{probability of alternation} was close to the
expected value in a random string,  participants tend to produce
$n$-grams of length $\geq 2$ with  probability distributions which
are not equiprobable (see Fig. \ref{fig:largeNumbersLaws}).
Strings containing more alternations (like $1010$, $0101$, $010$,
$101$) and $3-$ and $4-$ runs have a higher frequency than expected
by chance. This might be seen as an effort from participants to keep
the \textit{probability of alternation} close to $0.5$ by
compensating the excess of alternations with blocks of repetitions
of the same symbol.

\subsection{Comparing human randomness with other random sources}\label{subsec:comp}

We asked whether $K$, the Kolmogorov complexity relative to $\leng$
defined in section \ref{sec:model}\ref{subsec:KL}, is able to detect and
compress more patterns in strings generated by participants than in
strings produced by other sources, which are considered {\em random}
for many practical issues. In particular, we studied strings
originated by two sources: Pseudo-Random Number Generator (PRNG) and
Atmospheric Noise (AN).

%
%
%

For the PRNG source, we chose the Mersenne Twister algorithm
\cite{Matsumoto} (specifically, the second revision from 2002 that
is currently implemented in \textit{GNU Scientific Library}). The
atmospheric noise was taken from \textit{random.org} site
(property of Randomness and Integrity Services Limited) which also
runs real-time statistic tests recommended by the US National
Institute of Standards and Technology to ensure the random quality
of the numbers produced over time.

In Table \ref{tab:complexity}, we summarize our results using $b=1$
and $r=1$ for the parameters of $K$ as defined in
section \ref{sec:model}\ref{subsec:KL}

\begin{table}[h]
\centering
\begin{tabular}{lrrr}
 & {\em Participants} & {\em PRNG}  & {\em AN}  \\
\hline
\hline
Mean $\mu$ & 48.43 & 52.99 & 53.88    \\
Std $\sigma$  &  6.62 & 3.06 & 2.87     \\
$1^{st}$ quartile & 45.30 & 50.42 & 51.88     \\
Median       & 49.23 & 53.15 & 53.85 \\
$3^{rd}$ quartile  & 51.79 & 55.21 & 55.79 \\
\hline
\hline
\end{tabular}
\caption{Values of $K(x)$, where $x$ is a string produced by
participants, PRNG or AN sources} \label{tab:complexity}
\end{table}

The mean and median of $K$ increases when comparing participant's
string with PRNG or AN strings. This difference was significant, as
confirmed by a t-test  ($p$-value of $4.9 \times 10^{-11}$ when
comparing participant's sample with PRNG one, a $p$-value of $1.2
\times 10^{-15}$ when comparing participant's with AN and a
$p$-value of $1.4 \times 10^{-2}$ when comparing PRNG with AN
sample).

Therefore, despite the simplicity of $\leng$, based merely on
\textsl{prints} and \textsl{repeats}, it is rich enough to identify
regularities of human sequences. The $K$ function relative to
$\leng$ is   an effective and significant measure to distinguish
strings produced by participants with profound understanding in the
mathematics of randomness,  from PRNG and AN strings. As expected,
humans  produce less complex (i.e., less random) strings than those
produced by PRNG or atmospheric noise sources.

\subsection{Mental fatigue}

On cognitively demanding tasks, fatigue affects performance by
deteriorating the capacity to organize   behavior
\cite{Weiss64,fatigueBar,fatigueBro,fatigueFlo,fatigueHoc}.
Specifically, Weiss claim that boredom may be a factor that
increases non-randomness~\cite{Weiss64}. Hence, as another test to
the ability of $K$  relative to  $\leng$ to identify idiosyncratic
elements of human regularities, we asked whether the random quality
of the participant's string deteriorated with time.

For each of the $138$ strings $x=x_1\cdots x_{120}$
 ($x_i\in \{0,1\}$) produced by the participants,
we measured the $K$ complexity of all the sub-strings of
length~$30$.

\begin{figure}[htb]
\begin{tikzpicture}
\tikzstyle{every node}=[font=\small]
  
\begin{axis}[%
width=6.1cm,
height=3.1cm,
scale only axis,
xmin=1, xmax=91,
xlabel={$i$},
ymin=14, ymax=15,
ylabel={$K$},
axis lines*=left]
\addplot [
color=blue,
solid,
forget plot
]
coordinates{
 (1,14.3170054614604)(2,14.3299277850923)(3,14.27263038925)(4,14.3857236185067)(5,14.3799133636619)(6,14.3862321430805)(7,14.4525224513084)(8,14.4878506556124)(9,14.4344947000127)(10,14.562248374471)(11,14.5150572133744)(12,14.5307581405408)(13,14.5334811151447)(14,14.3380629162434)(15,14.3704422447247)(16,14.3317903342608)(17,14.3678093644232)(18,14.4251918642458)(19,14.3646149705615)(20,14.3120838974642)(21,14.3658701887698)(22,14.3872874903888)(23,14.4341449627869)(24,14.5043624639586)(25,14.3742976323918)(26,14.3599333480944)(27,14.4753694500078)(28,14.3596060429874)(29,14.2459340462634)(30,14.2770911830871)(31,14.3056104492873)(32,14.1691746552497)(33,14.0880306701488)(34,14.1545994418471)(35,14.1455114490937)(36,14.2407306005188)(37,14.2293698669754)(38,14.2284927792248)(39,14.2424514417286)(40,14.2680659222662)(41,14.28336827057)(42,14.306666159003)(43,14.3299988709583)(44,14.176223910985)(45,14.0793147001455)(46,14.1219622452182)(47,14.2506211450217)(48,14.2343610504164)(49,14.2127894837413)(50,14.1744012992228)(51,14.2164540325576)(52,14.2613618626889)(53,14.1466064824279)(54,14.1517784682615)(55,14.2244488945712)(56,14.1391613111301)(57,14.1423667631971)(58,14.1745171892554)(59,14.0404561314879)(60,14.1955520428061)(61,14.2044239925703)(62,14.1886985806762)(63,14.1210862392856)(64,14.1031863657237)(65,14.0688671421785)(66,14.1480928848526)(67,14.2313448147464)(68,14.3116427549335)(69,14.279807054689)(70,14.3196743695167)(71,14.2678099999574)(72,14.2099847086976)(73,14.0802209005385)(74,14.1174327751273)(75,14.1905206008464)(76,14.1871658740455)(77,14.2061281979175)(78,14.2164480059694)(79,14.0692070721305)(80,14.2348148399879)(81,14.2686224267285)(82,14.1893067175706)(83,14.2324455800091)(84,14.1839995164124)(85,14.2126394295259)(86,14.1420261815822)(87,14.0746091244718)(88,14.1472049134145)(89,14.2843934444665)(90,14.2458900893259)(91,14.2594215451466) 
};
\addplot [
color=green!50!black,
dashed,
line width=2.0pt,
forget plot
]
coordinates{
 (1,14.3170054614604)(2,14.3165180216078)(3,14.3160305817552)(4,14.3155431419026)(5,14.31505570205)(6,14.3145682621974)(7,14.3140808223447)(8,14.3135933824921)(9,14.3131059426395)(10,14.3126185027869)(11,14.3121310629343)(12,14.3116436230817)(13,14.3111561832291)(14,14.3106687433765)(15,14.3101813035239)(16,14.3096938636713)(17,14.3092064238186)(18,14.308718983966)(19,14.3082315441134)(20,14.3077441042608)(21,14.3072566644082)(22,14.3067692245556)(23,14.306281784703)(24,14.3057943448504)(25,14.3053069049978)(26,14.3048194651452)(27,14.3043320252925)(28,14.3038445854399)(29,14.3033571455873)(30,14.3028697057347)(31,14.3023822658821)(32,14.3018948260295)(33,14.3014073861769)(34,14.3009199463243)(35,14.3004325064717)(36,14.2999450666191)(37,14.2994576267664)(38,14.2989701869138)(39,14.2984827470612)(40,14.2979953072086)(41,14.297507867356)(42,14.2970204275034)(43,14.2965329876508)(44,14.2960455477982)(45,14.2955581079456)(46,14.295070668093)(47,14.2945832282403)(48,14.2940957883877)(49,14.2936083485351)(50,14.2931209086825)(51,14.2926334688299)(52,14.2921460289773)(53,14.2916585891247)(54,14.2911711492721)(55,14.2906837094195)(56,14.2901962695669)(57,14.2897088297142)(58,14.2892213898616)(59,14.288733950009)(60,14.2882465101564)(61,14.2877590703038)(62,14.2872716304512)(63,14.2867841905986)(64,14.286296750746)(65,14.2858093108934)(66,14.2853218710408)(67,14.2848344311881)(68,14.2843469913355)(69,14.2838595514829)(70,14.2833721116303)(71,14.2828846717777)(72,14.2823972319251)(73,14.2819097920725)(74,14.2814223522199)(75,14.2809349123673)(76,14.2804474725147)(77,14.279960032662)(78,14.2794725928094)(79,14.2789851529568)(80,14.2784977131042)(81,14.2780102732516)(82,14.277522833399)(83,14.2770353935464)(84,14.2765479536938)(85,14.2760605138412)(86,14.2755730739886)(87,14.2750856341359)(88,14.2745981942833)(89,14.2741107544307)(90,14.2736233145781)(91,14.2731358747255) 
};
\end{axis}
\end{tikzpicture}%
\caption{Average of $K(x_i \cdots x_{i+30})$ for
AN}\label{fig:fatigueAN}
\end{figure}
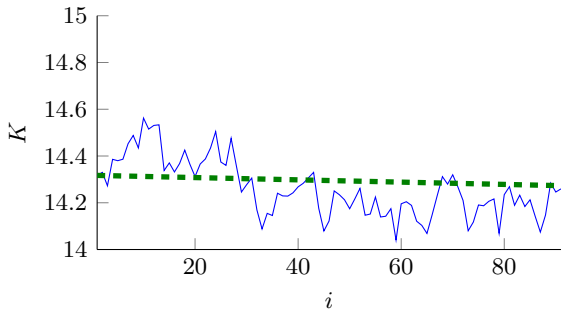

Specifically, we calculated the  average $K(x_i \cdots x_{i+30})$
from the $138$ strings for each $i \in [0,90]$ (see
Fig.~\ref{fig:fatigue}), using the same parameters as in
section \ref{sec:results}\ref{subsec:comp} ($b=r=1$), and compared to the
same sliding average procedure for PRNG
(Fig.~\ref{fig:fatiguePRNG}) and AN sources
(Fig.~\ref{fig:fatigueAN}).

\begin{figure}[htb]
\begin{tikzpicture}
\tikzstyle{every node}=[font=\small]

\begin{axis}[%
width=6.1cm,
height=3.1cm,
scale only axis,
xmin=1, xmax=91,
xlabel={$i$},
ymin=12.5, ymax=13.5,
ylabel={$K$},
axis lines*=left]
\addplot [
color=blue,
solid,
forget plot
]
coordinates{
(1,13.3597388803794)(2,13.4475306234157)(3,13.3550791229425)(4,13.2462996098636)(5,13.2326547512106)(6,13.4232010707248)(7,13.394810503035)(8,13.3782799965622)(9,13.3959580573255)(10,13.3013268194943)(11,13.3616192528511)(12,13.3235132111224)(13,13.1996473318673)(14,13.2473919804713)(15,13.2409121111882)(16,13.0934195343814)(17,13.0002810212778)(18,13.1416159004936)(19,13.0696727969746)(20,13.0265815170965)(21,13.1743270395822)(22,13.1507522818166)(23,13.0823120506297)(24,13.0621092108691)(25,13.0402888903623)(26,13.0891490884537)(27,13.0642269738149)(28,13.0833887676947)(29,13.1380827425402)(30,13.1729726115389)(31,13.074178599588)(32,13.1115371042788)(33,13.1454188835697)(34,13.1441358948838)(35,13.1189248139861)(36,13.119553479795)(37,13.1622927431444)(38,13.1051481494604)(39,12.9821946888563)(40,13.1092319547287)(41,13.2481324271977)(42,13.2335145783103)(43,13.2563532209133)(44,13.2307079316965)(45,13.3143246330251)(46,13.3324005941094)(47,13.1637132975756)(48,13.0953396270454)(49,13.2696462371936)(50,13.1567438444783)(51,13.0949705548667)(52,13.25072285731)(53,13.0992474594919)(54,13.0796266363625)(55,13.1147967381337)(56,13.1860275997644)(57,13.154940224007)(58,13.1779287034648)(59,13.113777277136)(60,12.9846484287926)(61,12.9268155359834)(62,12.9073376873446)(63,12.9537960066212)(64,13.1040327887254)(65,13.2052871912237)(66,13.1177817310756)(67,13.0531420995426)(68,13.0681391659289)(69,13.0231860869568)(70,12.9309434237653)(71,12.9485148175043)(72,12.8593572776616)(73,12.8697341121701)(74,12.7830160470474)(75,12.8153252766356)(76,12.8007294373214)(77,12.7746282727368)(78,12.7275944346229)(79,12.8433381926062)(80,12.811519571751)(81,12.9073410493321)(82,12.9771310082262)(83,12.8041291076016)(84,12.7092239811727)(85,12.7809235783977)(86,12.7786077054554)(87,12.6918203992999)(88,12.714529573857)(89,12.7606794170757)(90,12.7951773199877)(91,12.7157334710845)
};
\addplot [
color=green!50!black,
dashed,
line width=2.0pt,
forget plot
]
coordinates{
(1,13.3597388803794)(2,13.3527077654099)(3,13.3456766504403)(4,13.3386455354708)(5,13.3316144205012)(6,13.3245833055317)(7,13.3175521905621)(8,13.3105210755926)(9,13.3034899606231)(10,13.2964588456535)(11,13.289427730684)(12,13.2823966157144)(13,13.2753655007449)(14,13.2683343857754)(15,13.2613032708058)(16,13.2542721558363)(17,13.2472410408667)(18,13.2402099258972)(19,13.2331788109276)(20,13.2261476959581)(21,13.2191165809886)(22,13.212085466019)(23,13.2050543510495)(24,13.1980232360799)(25,13.1909921211104)(26,13.1839610061408)(27,13.1769298911713)(28,13.1698987762018)(29,13.1628676612322)(30,13.1558365462627)(31,13.1488054312931)(32,13.1417743163236)(33,13.134743201354)(34,13.1277120863845)(35,13.120680971415)(36,13.1136498564454)(37,13.1066187414759)(38,13.0995876265063)(39,13.0925565115368)(40,13.0855253965672)(41,13.0784942815977)(42,13.0714631666282)(43,13.0644320516586)(44,13.0574009366891)(45,13.0503698217195)(46,13.04333870675)(47,13.0363075917805)(48,13.0292764768109)(49,13.0222453618414)(50,13.0152142468718)(51,13.0081831319023)(52,13.0011520169327)(53,12.9941209019632)(54,12.9870897869937)(55,12.9800586720241)(56,12.9730275570546)(57,12.965996442085)(58,12.9589653271155)(59,12.9519342121459)(60,12.9449030971764)(61,12.9378719822069)(62,12.9308408672373)(63,12.9238097522678)(64,12.9167786372982)(65,12.9097475223287)(66,12.9027164073592)(67,12.8956852923896)(68,12.8886541774201)(69,12.8816230624505)(70,12.874591947481)(71,12.8675608325114)(72,12.8605297175419)(73,12.8534986025724)(74,12.8464674876028)(75,12.8394363726333)(76,12.8324052576637)(77,12.8253741426942)(78,12.8183430277246)(79,12.8113119127551)(80,12.8042807977856)(81,12.797249682816)(82,12.7902185678465)(83,12.7831874528769)(84,12.7761563379074)(85,12.7691252229378)(86,12.7620941079683)(87,12.7550629929988)(88,12.7480318780292)(89,12.7410007630597)(90,12.7339696480901)(91,12.7269385331206)
};
\end{axis}
\end{tikzpicture}%
\caption{Average of $K(x_i \cdots x_{i+30})$ for
participants}\label{fig:fatigue}
\end{figure}
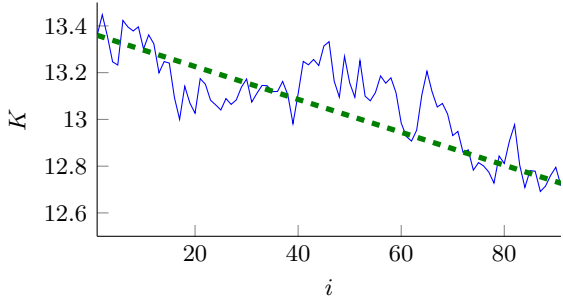

\begin{figure}[htb]
\begin{tikzpicture}
\tikzstyle{every node}=[font=\small]

\begin{axis}[%
width=6.1cm,
height=3.1cm,
scale only axis,
xmin=1, xmax=91,
xlabel={$i$},
ymin=13.5, ymax=14.5,
ylabel={$K$},
axis lines*=left]
\addplot [
color=blue,
solid,
forget plot
]
coordinates{
 (1,13.9383631762982)(2,13.8687250563818)(3,13.8472023313435)(4,13.8847304038252)(5,13.8698503664431)(6,13.9443810601278)(7,13.9497424969471)(8,13.999116473645)(9,14.0298275832387)(10,13.9547895236609)(11,13.9524914551708)(12,14.0658453023986)(13,14.0637910839418)(14,13.9713953402177)(15,13.9937871560762)(16,13.9946518524232)(17,14.1257584524539)(18,14.0245911920628)(19,14.1099888279852)(20,14.1651210233933)(21,14.2824514967935)(22,14.2111137581495)(23,14.2984736611837)(24,14.2137235925444)(25,14.1571570709938)(26,14.2404237384125)(27,14.2061994716731)(28,14.1315723000946)(29,14.2844624777235)(30,14.2847617466019)(31,14.1096293435566)(32,14.0996855586867)(33,14.1949922837751)(34,14.1819449933095)(35,14.2448205470759)(36,14.1497700825236)(37,14.1796895707544)(38,14.1514573098475)(39,14.0895548314995)(40,13.993141928348)(41,14.0346104779951)(42,14.0560898376878)(43,14.1413132503753)(44,14.1112910885489)(45,14.1308021545416)(46,14.0346203943115)(47,14.1452483070265)(48,14.1710666136674)(49,14.1307250061966)(50,14.1534247563184)(51,14.1215595343249)(52,14.1916813260221)(53,14.3267520426639)(54,14.2570574465877)(55,14.3310467115623)(56,14.2002405390905)(57,14.255675554878)(58,14.2104103529979)(59,14.0755316267568)(60,13.9894866538834)(61,13.9602661763308)(62,13.9618618705313)(63,14.002619718731)(64,14.0486773436514)(65,14.0354164053267)(66,14.0323605849967)(67,14.0687909424286)(68,14.1284043800787)(69,14.098040303439)(70,14.1539816855685)(71,14.0801496868539)(72,14.0877031549564)(73,14.0166321021667)(74,13.9641131690927)(75,13.9046828756996)(76,13.9286822911931)(77,13.8523679979135)(78,13.8207801629765)(79,13.7909381442031)(80,13.9237143606474)(81,14.0409141242234)(82,13.9135941077873)(83,13.9027958830985)(84,13.9985951952602)(85,14.0258866878889)(86,14.0044937431789)(87,14.0350356623022)(88,13.9623518959938)(89,14.1175128665897)(90,14.1143529713578)(91,14.119684496167) 
};
\addplot [
color=green!50!black,
dashed,
line width=2.0pt,
forget plot
]
coordinates{
 (1,13.9383631762982)(2,13.9400023853346)(3,13.9416415943711)(4,13.9432808034075)(5,13.9449200124439)(6,13.9465592214804)(7,13.9481984305168)(8,13.9498376395532)(9,13.9514768485896)(10,13.9531160576261)(11,13.9547552666625)(12,13.9563944756989)(13,13.9580336847354)(14,13.9596728937718)(15,13.9613121028082)(16,13.9629513118447)(17,13.9645905208811)(18,13.9662297299175)(19,13.967868938954)(20,13.9695081479904)(21,13.9711473570268)(22,13.9727865660632)(23,13.9744257750997)(24,13.9760649841361)(25,13.9777041931725)(26,13.979343402209)(27,13.9809826112454)(28,13.9826218202818)(29,13.9842610293183)(30,13.9859002383547)(31,13.9875394473911)(32,13.9891786564276)(33,13.990817865464)(34,13.9924570745004)(35,13.9940962835368)(36,13.9957354925733)(37,13.9973747016097)(38,13.9990139106461)(39,14.0006531196826)(40,14.002292328719)(41,14.0039315377554)(42,14.0055707467919)(43,14.0072099558283)(44,14.0088491648647)(45,14.0104883739012)(46,14.0121275829376)(47,14.013766791974)(48,14.0154060010104)(49,14.0170452100469)(50,14.0186844190833)(51,14.0203236281197)(52,14.0219628371562)(53,14.0236020461926)(54,14.025241255229)(55,14.0268804642655)(56,14.0285196733019)(57,14.0301588823383)(58,14.0317980913748)(59,14.0334373004112)(60,14.0350765094476)(61,14.036715718484)(62,14.0383549275205)(63,14.0399941365569)(64,14.0416333455933)(65,14.0432725546298)(66,14.0449117636662)(67,14.0465509727026)(68,14.0481901817391)(69,14.0498293907755)(70,14.0514685998119)(71,14.0531078088484)(72,14.0547470178848)(73,14.0563862269212)(74,14.0580254359576)(75,14.0596646449941)(76,14.0613038540305)(77,14.0629430630669)(78,14.0645822721034)(79,14.0662214811398)(80,14.0678606901762)(81,14.0694998992127)(82,14.0711391082491)(83,14.0727783172855)(84,14.074417526322)(85,14.0760567353584)(86,14.0776959443948)(87,14.0793351534312)(88,14.0809743624677)(89,14.0826135715041)(90,14.0842527805405)(91,14.085891989577) 
};
\end{axis}
\end{tikzpicture}%
\caption{Average of $K(x_i \cdots x_{i+30})$ for
PRNG}\label{fig:fatiguePRNG}
\end{figure}
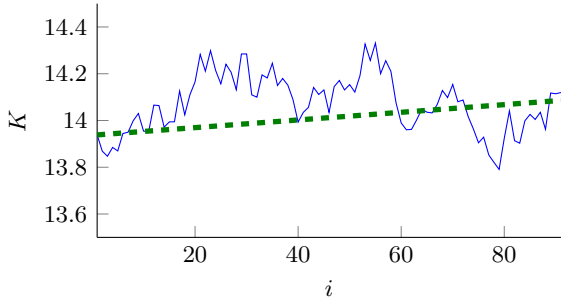

The sole source which showed a significant linear regression was
human generated data (see Table~\ref{tab:fatiguetest}) which, as
expected, showed a negative correlation indicating that participants
produced less complex or random strings over time (slope $-0.007$,
$p <0.02$).

\begin{table}[h]\scriptsize
\centering
\begin{tabular}{lrrr}
 & {\em Participants} & {\em PRNG}  & {\em AN}  \\
\hline
\hline
Mean slope & -0.007 & 0.0016 & -0.0005    \\
$p$-value  &  0.02 & 0.5 & 0.8     \\
CI & [-0.01,-0.001] & [-0.003,0.006] & [-0.005,0.004]     \\
\hline
\hline
\end{tabular}
\caption{Predictability} \label{tab:fatiguetest}
\end{table}
%



The finding of a fatigue-related effect shows that the
  unpropped, i.e.,\ resource-limited, human Turing machine is not
only limited in terms of the language it can parse, but also in terms
of the amount of time it can dedicate to a particular task.

\subsection{Predictability}

In section \ref{sec:model}\ref{subsec:predictor}, we introduced a
prediction method with two parameters: $\ell_F$ and $\ell_P$. A
predictor based on $\leng$ achieved levels of predictability close
to 56\% which were highly significant (see
Table~\ref{tab:basePredictability}). The predictor, as expected,
performed at chance for the control PRNG and AN data. This fit was
relatively insensitive to the values of $\ell_P$ and $\ell_F$,
contrary to our intuition that there may be a memory scale which
would correspond in this framework to a given length.

A very important aspect of this investigation, in line with the
prior work of~\cite{Griff03}, is to inquire whether specific
parameters are stable for a given individual. To this aim, we
optimized, for each participant, the parameters using the first 80
symbols of the sequence and then tested these parameters in the
second half of each segment (last 80 symbols of the sequence)

After this optimization procedure, mean predictability increased
significantly to  $58.14\%$ ($p < 0.002$, see
Table~\ref{tab:optimizedPredictability}). As expected, the
optimization based on partial data of PRNG and AN resulted in no
improvement in the classifier, which remained at chance with no significant difference ($p < 0.3$, $p < 0.2$, respectively).

Hence, while the specific parameters for compression vary widely across each individual,
they show stability in the time-scale of this experiment.

\begin{table}[h]\small\centering
\begin{tabular}{lrrr}
 & {\em Participants} & {\em PRNG}  & {\em AN}  \\
\hline
\hline
Mean $\mu$ & 56.16 & 50.69 & 49.48    \\
Std $\sigma$  &  0.07 & 0.02 & 0.02     \\
$1^{st}$ quartile & 49.97 & 48.84 & 48.30     \\
Median       & 55.02 & 50.77 & 49.04 \\
$3^{rd}$ quartile  & 59.75 & 52.21 & 50.46 \\
\hline
\hline
\end{tabular}
\caption{Average predictability} \label{tab:basePredictability}

\bigskip

\begin{tabular}{lrrr}
 & {\em Participants} & {\em PRNG}  & {\em AN}  \\
\hline
\hline
Mean $\mu$ & 58.14 & 51.20 & 49.01    \\
Std $\sigma$  &  0.07 & 0.04 & 0.03     \\
$1^{st}$ quartile & 52.88 & 48.56 & 47.11     \\
Median       & 56.73 & 50.72 & 49.28 \\
$3^{rd}$ quartile  & 62.02 & 53.85 & 50.48 \\
\hline
\hline
\end{tabular}
\caption{Optimized predictability}
\label{tab:optimizedPredictability}
\end{table}



\section{Discussion}\label{sec:discussion}


Here we analyzed strings produced by participants attempting to
generate random strings. Participants had a profound understanding
of randomness and hence avoided typical misconceptions such as
exaggerating the number of alternations. We reasoned that remaining
regularities would express the algorithmic nature of human thoughts,
revealed in the form of specific patterns.

Our effort here was to
bridge the gap between Kolmogorov theory and psychology, developing a concrete language, $\leng$, satisfying the following requirements: 1)  to be simple enough so that
the complexity of any given sequence can be computed, 2)  to be based
on tangible operations of human reasoning (\textsl{ printing},
\textsl{ repeating}, \dots), 3)  to be sufficiently powerful to generate
all possible sequences but not too powerful as to identify
regularities which would be {\em invisible} to humans.

More specifically, our aim is to develop a class of languages with
certain degrees of freedom which can then be fit to an individual
(or an individual in a specific context and time). Here, we opted for
a comparably easier strategy by only allowing the relative cost of
each operation to vary. However, a natural extension of this
framework is to generate classes of languages where structural and
qualitative aspects of the language
are free to vary. For instance, one can devise a program structure
for repeating portions of (not necessarily neighboring) code, or
considering the more general framework of {\em for-programs} where
the repetitions are more general than in our setting: \textsl{for
i=1 to n do $P(i)$}, where $P$ is a program that uses the successive
values of $i=1,2,\dots,n$ in each iteration. For instance, the
following program
\begin{eqnarray*}
&&\small \textsl{for i=1 to 6 do }\\
&&\small \qquad\textsl{print `0'}\\
&&\small \qquad\textsl{repeat i times: print `1'}
\end{eqnarray*}
would describe the string
\begin{center}
\small $010110111011110111110111111$.
\end{center}
The challenge from the computational theoretical point of view is to
define an extension which induces a computable (even more, feasible,
whenever possible) Kolmogorov complexity. For instance, adding simple
control structures like conditional jumps or allowing the use of
imperative program variables may turn the language into
Turing-complete, with the theoretical consequences that we already
mentioned. The aim is to keep the language simple and yet include
structures to compact some patterns which are compatible with the
human language of thought.

We emphasize that our aim here was not to generate an optimal
predictor of human sequences. Clearly, restricting $\leng$ to a very
rudimentary language is not the way to go to identify vast classes
of patterns. Our goal, instead, was to use human sequences to
calibrate a language which expresses and captures specific patterns
of human thought in a tangible and concrete way.


Our model is based on ideas from Kolmogorov complexity and
Solomonoff's induction. It is important to  compare it to what we
think is the closest and more similar approach in previous studies:
the work~\cite{Griff03} of Griffiths and Tenenbaum's. Griffiths and
Tenenbaum devise a series of statistical models that account
for different kind of regularities. Each  model $Z$ is fixed and
assigns to every binary string $x$ a probability $P_Z(x)$. This
probabilistic approach is connected to Kolmogorov complexity theory
via Levin's famous Coding Theorem, which points out a remarkably
numerical relation between the algorithmic probability $P_U(x)$ (the
probability that the universal prefix Turing machine $U$ outputs $x$
when the input is filled-up with the results of coin tosses) and the
(prefix) Kolmogorov complexity $K_U$ described in
section \ref{sec:intro}\ref{sec:kolmogorov} Formally, the theorem states that there is a constant $c$ such that for any string $x\in\{0,1\}^*$ such that

\begin{equation}\label{eqn:coding-theorem}
 |-\log P_U(x)-K_U(x)|\leq c
\end{equation} (the reader is referred to section 4.3.4 of~\cite{LV08} for
more details). Griffiths \& Tenenbaum's bridge to Kolmogorov
complexity is only established through this last theoretical result:
replacing $P_U$ by $P_Z$ in Eq. (\ref{eqn:coding-theorem})
should automatically give us {\em some} Kolmogorov complexity $K_Z$
with respect to {\em some} underlying Turing machine $Z$.

While there is hence a formal relation to  Kolmogorov complexity,  there
is no explicit definition of the underlying {\em machine}, and hence no notion of {\em program}.

On the contrary, we propose a specific {\em language of thought},
formalized as the programming language $\leng$ or, alternatively, as
a Turing machine $\M$, which assigns formal {\em semantics} to each
program. Semantics are given, precisely, through the behavior of
$\M$. The fundamental introduction of program semantics and the
clear distinction between {\em inputs} (programs of $\M$) and {\em
outputs} (binary strings) allows us to give a straightforward
definition of Kolmogorov complexity relative to $\M$, denoted
$K_{\M}$, which ---because of the choice of $\leng$--- becomes
computable in polynomial time. Once we count with a complexity
function, we apply Solomonoff's ideas of inductive inference to
obtain a predictor which tries to guess the continuation of a given
string under the assumption that the most probable one is the most
compressible in terms of $\leng$-Kolmogorov complexity. As
in~\cite{Griff03}, we also make use of the Coding
Theorem~(\ref{eqn:coding-theorem}), but in the opposite direction:
given the complexity $K_\M$, we derive an algorithmic probability
$P_\M$.

This work is mainly a theoretical development, to develop a
framework to adapt Kolmogorov ideas in a constructive procedure
(i.e., defining an explicit language) to identify regularities in
human sequences. The theory was validated experimentally, as three
tests were satisfied: 1) human sequences were less complex than
control PRNG sequences, 2) human sequences were  non-stationary,
showing decreasing values of complexity, 3) each individual showed
traces of algorithmic stability since fitting of partial data was
more effective to predict subsequent data than average fits. Our
hope is that this theory may constitute, in the future, a useful
framework  to ground and describe the patterns of human thoughts.

\begin{acknowledgements}
The authors are thankful to Daniel Gor\'{\i}n and Guillermo Cecchi for useful
discussions. S.\ Figueira is partially supported by grants
PICT-2011-0365 and UBACyT 20020110100025.
\end{acknowledgements}


\end{document}